\documentclass[useAMS,usenatbib]{mn2e}
\usepackage{psfig}

\def\kms{\relax \ifmmode {\,\rm km\,s}^{-1}\else \,km\,s$^{-1}$\fi}

\def\mincir{\ \raise-2.truept\hbox{\rlap{\hbox{$\sim$}}\raise5.truept
    \hbox{$<$}\ }}
\def\magcir{\ \raise-2.truept\hbox{\rlap{\hbox{$\sim$}}\raise5.truept
    \hbox{$>$}\ }}

\def\arcsec{\hbox{$^{\prime\prime}$}}
\def\nii{[N {\sc ii}]}
\def\sii{[S {\sc ii}]}

\def\oii{[O {\sc ii}]}

\def\heii{He{\sc ii}}
\def\hei{He{\sc i}}
\def\oiii{[O {\sc iii}]}

\def\cliii{[Cl {\sc iii}]}

\def\hb{H$\beta$}
\def\ki{$\rm{Knot1}$}

\def\chb{$c_{\rm H\beta}$}
\def\te{T$_e$}
\def\teff{T$_{eff}$}
\def\ne{N$_e$}
\def\pi{\rm {Paper~{\sc i}}}
\def\an{\AA}

\title[Nitrogen abundance of the NGC~7009's outer knots]{On the nitrogen abundance of FLIERs: the outer 
knots of the planetary nebula NGC~7009}
\author[Gon\c calves, Ercolano, et al. ]{D. R. Gon\c calves$^{1,3}$\thanks{E-mail:
denise@astro.iag.usp.br (DRG)} \& B. Ercolano$^{2}$; \newauthor  
and A. Carnero$^{2}$;  A. Mampaso$^{3}$; R. L. M. Corradi$^{3,4}$\\
$^{1}$IAG, Universidade de S\~ao Paulo, Rua do Mat\~ao 1226, 05508-900, 
S\~ao Paulo, Brazil\\
$^{2}$Department of Physics and Astronomy, University College London, Gower Street, 
WC1E 6BT, London, UK\\
$^{3}$Instituto de Astrof\'{\i}sica de Canarias, E-38205 La Laguna,
Tenerife, Spain\\
$^{4}$Isaac Newton Group of Telescopes, Apartado de Correos 321, E-38700 Sta.Cruz de 
La Palma, Spain\\}
\begin{document}

\date{Accepted ??. Received ??; in original form ??}

\pagerange{\pageref{firstpage}--\pageref{lastpage}} \pubyear{2005}

\maketitle

\label{firstpage}

\begin{abstract}
We have constructed a 3D photoionisation model of a planetary nebula (PN) similar in structure to NGC~7009 
with its outer pair of knots (also known as FLIERs --fast, low-ionization emission regions). The work is 
motivated by the fact that the strong \nii$\lambda$6583 line emission from FLIERs in many planetary 
nebulae has been attributed to a significant local overabundance of nitrogen. We explore the 
possibility that the apparent enhanced nitrogen abundance  previously reported in the FLIERs may be due to 
ionization effects. The model is constrained by the results obtained by Gon\c calves, Corradi, Mampaso 
\& Perinotto (2003) from the analysis of both HST \oiii\ and \nii\ images, and long-slit spectra of NGC~7009.  
Our model is indeed able to reproduce the main spectroscopic and imaging characteristics 
of NGC~7009's bright inner rim and its outer pairs of knots, assuming {\it homogeneous elemental abundances} 
throughout the nebula, for {\it nitrogen} as well as all the other elements included in the model.

We also study the effects of a narrow slit on our non-spherically symmetric density distribution, via the 
convolution of the model results with the profile of the long-slit used to obtain the spectroscopic observations 
that constrained our model. This effect significantly enhances the \nii/\hb\ emission, more in the FLIERs than in 
the inner rim.

Because of the fact that the (N$^+$/N)/(O$^+$/O) ratio predicted by our models are 0.60 for the rim and 0.72 for 
the knots, so clearly in disagreement with the N$^+$/N=O$^+$/O assumption of the ionization correction factors 
method ({\it icf}), the {\it icf}s will be underestimated by the empirical scheme, in both components, rim and knots, 
but more so in the knots. This effect is partly responsible for the apparent inhomogeneous N abundance empirically 
derived. The differences in the above ratio in these two components of the nebula may be due to a number of 
effects including charge exchange --as pointed out previously by other authors-- and the difference in the 
ionization potentials of the relevant species --which makes this ratio extremely sensitive to the shape of 
the local radiation field. Because of the latter, a realistic density distribution is essential to the modelling 
of a non-spherical object, if useful information is to be extracted from spatially resolved observations, 
as in the case of NGC~7009. 

\end{abstract}

\begin{keywords}
Atomic data - ISM: abundances - planetary nebulae: individual (NGC~7009).
\end{keywords}

\section{Introduction}

\begin{table*}
\begin{center}
 \caption{Main physical/chemical parameters from \pi.} 
\begin{tabular}{lrrrrr}
\hline
\hline
\noalign{\smallskip}
Parameter & \multicolumn{2}{c}{R} & \multicolumn{2}{c}{K} & \multicolumn{1}{c}{NEB}\\
       & Eastern     & Western  & Eastern     & Western & Whole Nebula    \\
\noalign{\smallskip}
\hline
\multicolumn{6}{l}{}\\
\ne\sii (cm$^{-3}$)  & 5,500     & 5,900     & 2,000	& 1,300    & 4,000    \\
\ne\cliii (cm$^{-3}$)& 5,200     & 5,900     & -	& 1,900    & 1,300    \\
\te\oiii (K)         & 10,000    & 10,200    &  9,600	& 10,400   & 10,100   \\
\te\nii (K)          & 10,400    & 12,800    & 11,000	& 11,700   & 10,300   \\
\te\sii (K)          & -         & -         &  7,100	&  9,400   & -        \\
He/H                 & 1.08(-1)  & 1.16(-1)  & 1.02(-1) & 9.55(-2) & 1.11(-1) \\
O/H                  & 4.5(-4)   & 4.82(-4)  & 5.8(-4)  & 4.5(-4)  & 4.71(-4) \\ 
N/H                  & 7.0(-5)   & 1.8(-4)   & 3.8(-4)  & 2.5(-4)  & 1.7(-4)  \\
Ne/H                 & 1.1(-4)   & 1.1(-4)   & 1.1(-4)  & 1.3(-4)  & 1.1(-4)  \\
S/H                  & 6.1(-6)   & 4.9(-6)   & 1.39(-5) & 9.3(-6)  & 8.3(-6)  \\
\noalign{\smallskip}
\hline
\hline
\end{tabular}
\end{center}
\small{Empirically derived parameters, for the different structures of the 
NGC~7009, for both, the Eastern and Western sides of the nebula, with ``R",
``K" and ``NEB" standing for rim, outer knots and the whole nebula (see Tables 1 
and 3 of  \pi). $\ \ \ \ \ \ \ \ \ \ \ \ \ \ \ \ \ \ \ \ \ \ \ \ \ \ \ \ \ \ \ \ \ $} 
\end{table*}

It is well known that planetary nebulae (PNe) possess a number of 
small-scale structures that, at variance with their large-scale 
structures, are prominent in low-ionization 
emission lines, such as \nii\ and \oii. In terms of morphology and 
kinematics, the low-ionization structures 
\citep[LISs, for a review see ][]{b018} may appear as knots, filaments, jets, 
and isolated features that, in some cases, move with the same velocity of the 
ambient nebula, or supersonically through the environment, like FLIERs \citep[fast, 
low-ionization emission regions,][]{b05} or BRETs \citep[bipolar, rotating, 
episodic jets;][]{b028}.  

For over ten years, the strong \nii\ line emission from LISs has been 
attributed to a significant local overabundance of nitrogen (Balick et al.~1993,  
1994, 1998; and references therein). \citet{b05} 
interpreted the N-enrichment in FLIERs as evidence of their origins in recent 
high-velocity ejections of the PN central star. The above work and successively \citet{b06}, 
which included the derivation of the ionic and elemental abundances of LISs, pointed out 
``an apparent enhancement of nitrogen relative to hydrogen by factors of 2-5" in FLIERs. 
However, since then, further work casted doubts over the latter statement \citep{b021,b01,b017,b033}. 

Abundances in PNe can be derived from the analysis of collisionally excited 
line (CEL) or optical recombination 
line (ORL) spectra using empirical methods or tailored photoionisation models 
\citep[e.g.][]{b040} or a combination of the two. For some elements (e.g. He) and ions 
(e.g. N$^{++}$ if only the optical spectrum is available) the use of ORL is mandatory, 
whilst for others the use of CELs may be unavoidable. But, when both CEL
and ORL determinations are available, the former have
been commonly preferred because they are generally stronger and easier
to detect than ORLs. Moreover, 
there is a well known discrepancy between CEL- and ORL-abundances, as well as 
electron temperature determinations \citep{b027}. 
Empirical abundance analysis rely on ionization correction factors ({\it icf}s) 
to account for the unseen ions (e.g. Kingsburgh \& Barlow, 1994). 
Results obtained with the {\it icf} method can be somewhat uncertain in some 
cases, particularly when they are applied to spatially resolved long-slit 
spectra \citep{b01}, as has been the case for the work of \citet{b06,b021} 
and \citet[][hereafter \pi]{b017}.

In fact, the abundances derived in \pi\ and shown here in Table~1, from optical 
long-slit spectra of NGC~7009, 
using the {\it icf} scheme of \citet{b025}, showed only a marginal evidence for overabundance of N/H in the 
outer knots of the nebula (the ansae), reinforcing the doubts over previous results \citep{b06} 
where the N/H enhancement of a factor of 2-5 in the ansae were reported.

One of the major shortcomings of empirically-determined chemical abundances lies in the 
fact that a number of assumptions on the ionization structure of the gas need to be made in order to obtain 
the {\it icf}s. 
A preferred alternative would be the construction of a tailored photoinisation model for a given 
object, aiming to fit the emission line spectrum and, in the case of spatially resolved objects, projected 
maps in a number of emission lines. 

NGC~7009, the ``Saturn Nebula'', is a PN comprising a bright elliptical rim. Its small-scale 
structures include a pair of jets and two pairs of low-ionization knots.  On a larger scale, it is known 
that NGC~7009 possess a tenuous halo with a diameter of more than 4 arcmin \citep{b0032}, whose inner regions 
display a system of concentric rings \citep{b009} like those observed in NGC 6543 and other few PNe \citep{b07}.
High-excitation lines dominate the inner 
regions along the minor axis, while emission from low-ionization species is enhanced at the extremities of the 
major axis. The ionization structure is further enriched by the fact that the low-excitation regions present 
strong variations in excitation level and clumpiness. NGC~7009 was classified as an oxygen-rich PN \citep{b022}, 
with an O/C ratio exceeding 1, and anomalous N, O, and C abundances \citep{b02,b06,b022}. Its central star is an 
H-rich O-type star, with effective temperature of 82\,000~K \citep{b032,b024}.  The kinematics of NGC~7009 was 
studied first by \citet{b037} and \citet{b04}, who showed that the ansae are expanding near the plane of the sky 
at highly supersonic velocities. The derived inclination of the inner (caps) and outer (ansae) knots, with respect 
to the line of sight, are $i\cong 51^{\circ}$ and $i\cong 84^{\circ}$, respectively \citep{b037}.  More recently, 
\citet{b014} have measured the proper motion and kinematics of the ansae in NGC~7009, assuming that they are equal 
and opposite from the central star, obtaining V$_{\rm {exp}}$ = 114 $\pm$ 32 \kms, for the distance of 
$\sim$ 0.86 $\pm$ 0.34 kpc.

In this paper we will present a simple 3D photoionisation model, aiming at reproducing the observed geometry and 
spectroscopic ``peculiarities'' of a PN like NGC~7009, exploring the possibility that the enhanced \nii\ emission 
observed in the outer knots may be due to ionization effects. We will use the 3D photoinisation code {\sc mocassin} 
\citep{b010} together with the long-slit data presented in \pi\ and the {\it HST} images available for this nebula. 
Our model and the observational data are described in Section~2. The results for our main model are presented 
and discussed in Section~3, while in Section~4 we discuss an {\it alternative model}, aimed at highlighting the 
relevance of the geometry and density distribution in such a kind of modelling. The work is briefly summarised 
in Section~5, where our final conclusions are also stated.  

\section{Method}

\subsection{Observational data}

The observations used to constrain the photoionisation model were described in detail in \pi. These included {\it HST} 
\oiii\ and \nii\ images as well as Isaac Newton Telescope long-slit, intermediate dispersion spectra, 
along the PN major axis (P.A. = 79$^{\circ}$). For further details see \pi. 

We refer to Figure~1 of \pi, the {\it HST} \oiii\ and \nii\ images of NGC~7009, in which the nebular features 
are outlined. Here we will employ the same nomenclature introduced in Paper~{\sc i}. See also the left panel of Figure~2, 
in Section~3.5.

\subsection{The 3D photoionisation code: {\sc mocassin}}

The nebula was modeled using the 3D photoinisation code, {\sc mocassin}, of
\citet{b010}. The code employs a Monte Carlo technique to the solution of the radiative transfer of the stellar and 
diffuse field, allowing a completely geometry-independent treatment of the problem without the need of imposing 
symmetries or approximations for the transfer of the diffuse component. 

The reliability of the code was demonstrated via a set of benchmarks described
by \citet{b035} and \citet{b010}. A number of axy-symmetric planetary nebulae have 
already been modeled using {\sc mocassin}, examples include NGC~3918
\citep{b011}, NGC~1501 \citep{b013} and the H-deficient knots of Abell~30
\citep{b012}. 

\subsection{Input parameters}
\label{subsec:input}

\begin{table}
\label{tab:input}
\begin{center}
\caption{Input parameters for the model.}
\begin{tabular}{llll}
\hline
\hline
\multicolumn{4}{l}{} \\           
L$_*$ ($L_{\odot}$) & 3136        & N/H    & 2.0(-4) \\
T$_{eff}$ (K)       & 80,000 	  & O/H    & 4.5(-4) \\
R$_{in}$ (cm)       & 0.0 	  & Ne/H   & 1.06(-4) \\
R$_{out}$ (cm)      & 3.88(17)    & S/H    & 0.9(-5)  \\
He/H                & 0.112       & Ar/H   & 1.2(-6)  \\
C/H                 & 3.2(-4)     & Fe/H   & 5.0(-7)  \\
\hline
\hline
\end{tabular}
\end{center}
\small{Abundances are given by number, relative to H.} 
\end{table}

A thorough investigation of the vast parameter space was carried out in this work. This involved experimenting with 
various gas density distributions, central star parameters and nebular elemental abundances. The model input parameters 
that best fitted all observational constraints are summarised in Table~\ref{tab:input}, and discussed in more detail 
in the following subsections. 

A common problem when studying galactic PNe is the large uncertainties associated with the distance estimates, that 
propagate to the determination of the nebula geometry and central star parameters. We adopted a distance of 0.86~kpc 
for our models of NGC~7009, as computed by \citet{b014} from the weighted average of 14 values \citep{b00} determined 
with statistical methods; the value was quoted with an error of $\pm$ 0.34~kpc.

\subsubsection{Stellar parameters}

After having experimented with various stellar atmosphere models to describe the ionising continuum, we reverted to 
using a blackbody of \teff~=~80,000\,K and $\log$~L$_*$~=~3.50, as this resulted in the best fit of the nebular 
emission line spectrum. \citet{b032} determined the \teff\ and $\log$~g for the central star of NGC~7009 using non-LTE 
model atmosphere analysis of the stellar H and He absorption line profiles, finding \teff\ = 82 000~K and $\log$~L$_*$ 
= 3.97, in solar units. \citet{b032} assumed a distance of 2.1~kpc in their analysis which therefore resulted in their 
value for the stellar luminosity being higher than that inferred from our modelling. 

\subsubsection{Elemental Abundances}

The nebular elemental abundances used for the photoionisation model are listed 
in Table~2, where they are given by number with respect to H. Although the 
{\sc mocassin} code can handle chemical inhomogeneities, these were not 
included in our models as they proved to be not necessary to reproduce the 
CEL spectra of the R and K regions of NGC~7009. 

The values shown result from an iterative process, where the initial guesses at the elemental abundances of He, N, O, 
Ne and S, taken from \pi\ (see Table~1), and  those of C and Ar, from 
\citet{b036}, were successively modified to fit the spectroscopic observations. 

\subsubsection{Density distribution}

The simplest possible density distribution model was constructed in order to demonstrate that the spectroscopic 
peculiarities often found in LISs can be the product of simple (and well-known) photoionisation effects. The case 
of NGC~7009 is 
taken as an example, but the emphasis is not in the construction of a detailed model for this object in particular. 
With this in mind we described the nebula by an ellipsoidal rim with a H number density, N$_H$, peaking to 9000~cm$^{-3}$ 
in the short axis direction exponentially decreasing to a minimum value of 4000~cm$^{-3}$ in the long axis direction. The  
short and long axes of the inner and outer ellipsoids measure 3.84$\times$10$^{16}$~cm and 9.99$\times$10$^{16}$~cm, and 
7.06$\times$10$^{16}$~cm and 1.84$\times$10$^{17}$~cm, respectively, at the distance assumed for NGC~7009. The rim is 
surrounded by a spherical shell of less opaque, homogeneous density gas, with N$_H$~=~1600\,cm$^{-3}$. The diameter of 
the sphere is equal to the long axis of the outer ellipsoid defining the rim. Cylindrical jets, 1.75$\times$10$^{16}$~cm 
in diameter, connect the rim to a pair of disk-shaped knots aligned at a distance of 
3.49$\times$10$^{17}$~cm from the central star, along the long axis of the ellipsoid. The cylindrical jets widen into 
cone-shapes at the knot ends in order to simulate the effect of material accumulating at the knots, as suggested by the 
$HST$ images (particularly for knot K4, as seen in the right panels of Figure~2). The diameter of the base of the cones 
equals that of the disk-shaped knots. The centres of the 3.49$\times$10$^{16}$~cm diameter circular disks representing 
the knots are aligned with the centres of the cylindrical jets (hence they are seen almost edge on). The width of the 
disks is assumed to be 3.88$\times$10$^{16}$~cm in our model, although only a fraction of these is ionised, as is clear 
from the right panels of Figure~2. The H number density in the jets and knots is taken to be homogeneous and equal to 
1250~cm$^{-3}$ and 1500~cm$^{-3}$, respectively, consistently with the values derived in \pi. 

Results from the ellipsoidal rim and the spherical outer shell are combined into a single {\it R-component} to enable us 
to carry out a direct comparison with the slit spectra from \pi, as we show in
Table~1.

The jets ({\it J-component}) are included in our simulation as the radiation field has 
to be transferred through them before reaching the outer ansae ({\it K-component}). 
However, given that the emission detected from this region is very faint and 
that such structures may not be in equilibrium, we take our results for the 
{\it J-component} as very uncertain and omit them from any further discussion. 

Finally, neither the inner caps nor the tenuous halo were included in our model. 
The former because they do not lie on the same axis as the {\it K-component} 
\citep{b037}, and are therefore not expected to have a major influence on the 
ionization structure of the outer knots. And the latter because it is very faint, 
and therefore it is not expected to contribute significantly to the integrated 
emission line spectrum.

Figure~1 shows model profiles along the major axis, in agreement with our 
assumed density, as N$_e$/N$_H$ ratio is correlated to the level of ionisation 
of the gas (see Section~3.4).

\begin{table*}
\begin{center}
  \caption{Model and observed dereddened spectra of R, K and NEB. Line intensities are normalized to \hb=100.} 
\begin{tabular}{llllllllll}
\hline
\hline
\noalign{\smallskip}
Line Identification (\an) & \multicolumn{3}{c}{R} & \multicolumn{3}{c}{K} & \multicolumn{3}{c}{NEB}\\
       & Model     & Model  & Obs. &  Model	 & Model  & Obs. & Model	 & Model  & Obs.  \\
       & no-slit & slit &      & no-slit & slit &      & no-slit & slit	&       \\
\noalign{\smallskip}
\hline
\multicolumn{10}{l}{}\\
\hb (10$^{-13}$ erg cm$^{-2}$ s$^{-1}$)& 3119 & 397.1 & - &  7.93 & 2.94 & - & 3136 & 405.7 & 3197* \\ \\
\hline
{}[O{\sc ii}] 3726.0 + 3728.8 & 5.38 & 5.88 & 15.5 &  213. & 241. & 204. & 6.06 & 8.21 & 24.2\\
                              &      &      & 7.34 & 	   &	  & 155. &	&      &     \\
{}[Ne{\sc iii}] 3868.7        & 111. & 112. & 106. &  131. & 129. & 82.3 & 111. & 113. & 108.\\
                              &      &      & 105. & 	   &	  & 141. &	&      &     \\
{}**[Ne{\sc iii}] 3967.5       & 34.5 & 34.8 & 51.7 &  40.5 & 40.1 & 35.3 & 34.6 & 34.9 & 48.9\\
                              &      &      & 44.6 & 	   &	  & 59.8 &	&      &     \\
{}[S{\sc ii}] 4068.6          & 0.31 & 0.35 & 1.50 &  4.95 & 5.46 & 7.43 & 0.33 & 0.40 & 1.98\\
                              &      &      & 2.36 & 	   &	  & 5.04 &	&      &     \\
{}[S{\sc ii}] 4076.4          & 0.10 & 0.11 & 0.87 &  1.63 & 1.80 & 2.36 & 0.10 & 0.13 & 1.11\\
                              &      &      & 1.16 & 	   &	  & 1.83 &	&      &     \\
{}[O{\sc iii}] 4363.2         & 8.54 & 8.55 & 7.76 &  10.2 & 9.88 & 6.94 & 8.55 & 8.59 & 8.15\\
                              &      &      & 8.65 & 	   &	  & 9.64 &	&      &     \\
He{\sc ii} 4685.7             & 16.8 & 22.0 & 26.0 &  0.00 & 0.00 & 1.00 & 16.7 & 21.6 & 15.8\\ 
                              &      &      & 23.4 & 	   &	  & 1.23 &	&      &     \\
{}**[Ar{\sc iv}] 4711          & 3.92 & 3.87 & 5.48 &  1.50 & 1.40 & 2.19 & 3.91 & 3.84 & 4.10\\
                              &      &      & 4.96 & 	   &	  & 2.00 &	&      &     \\
{}[Ar{\sc iv}] 4740.2         & 5.19 & 5.24 & 5.56 &  1.27 & 1.19 & 1.14 & 5.17 & 5.18 & 3.92\\
                              &      &      & 4.54 & 	   &	  & 1.49 &	&      &     \\
H$\beta$ 4861.3	              & 100. & 100. & 100. &  100. & 100. & 100. & 100. & 100. & 100.\\
                              &      &      & 100. & 	   &	  & 100. &	&      &     \\
{}[O{\sc iii}] 5006.8         & 1197 & 1172 & 1162 &  1315 & 1272 & 1238 & 1198 & 1177 & 1206\\
                              &      &      & 1225 & 	   &	  & 1310 &	&      &     \\
{}[Cl{\sc iii}] 5517.7        & 0.35 & 0.36 & 0.43 &  1.09 & 1.08 & 0.00 & 0.35 & 0.37 & 0.54\\
                              &      &      & 0.43 & 	   &	  & 0.97 &	&      &     \\
{}[Cl{\sc iii}] 5537.9        & 0.57 & 0.59 & 0.53 &  1.04 & 1.03 & 0.00 & 0.57 & 0.59 & 0.64\\
                              &      &      & 0.55 & 	   &	  & 0.90 &	&      &     \\
{}[N{\sc ii}] 5754.6          & 0.10 & 0.11 & 0.14 &  3.04 & 3.48 & 6.49 & 0.11 & 0.15 & 0.46\\
                              &      &      & 0.18 & 	   &	  & 3.91 &	&      &     \\
He{\sc i} 5875.7              & 14.1 & 13.4 & 13.9 &  15.8 & 15.9 & 18.8 & 14.1 & 13.5 & 14.5\\
                              &      &      & 14.1 & 	   &	  & 15.0 &	&      &     \\
{}[S{\sc iii}] 6312.1         & 1.47 & 1.55 & 1.27 &  4.04 & 3.99 & 3.89 & 1.48 & 1.59 & 1.68\\
                              &      &      & 1.14 & 	   &	  & 3.25 &	&      &     \\
{}[N{\sc ii}] 6583.4          & 5.87 & 6.64 & 7.09 &  167. & 193. & 355. & 6.39 & 8.44 & 27.\\
                              &      &      & 5.60 & 	   &	  & 194. &	&      &     \\
He{\sc i} 6678.1              & 3.99 & 3.82 & 3.87 &  4.48 & 4.49 & 7.49 & 3.99 & 3.83 & 3.97\\
                              &      &      & 3.95 & 	   &	  & 3.18 &	&      &     \\
{}[S{\sc ii}] 6716.5          & 0.46 & 0.53 & 0.48 &  21.9 & 24.2 & 36.8 & 0.53 & 0.79 & 2.33\\
                              &      &      & 0.38 & 	   &	  & 23.2 &	&      &     \\
{}[S{\sc ii}] 6730.8          & 0.84 & 0.99 & 0.84 &  29.3 & 32.4 & 51.0 & 0.94 & 1.32 & 3.85\\
                              &      &      & 0.69 & 	   &	  & 28.4 &	&      &     \\
\noalign{\smallskip}
\hline
\hline
\end{tabular}
\end{center}
\small{The Model spectra are given with and without considering the narrow slit effect. Upper rows 
give the observed intensities of the North-East (R1 and K1) part of the nebula, while 
those for the South-West (R2 and K4) zone are given in the lower rows. See Figure~1 of \pi\ for the sizes of 
R1, K1, R2 and K4. ``0.00" as the model predicted \heii$\lambda$4686 emission  and as the observed 
[Cl{\sc iii}]$\lambda\lambda$5517,5737 from K, former means that model does not produce any such emission and 
latter means that such lines were not detected in the spectra of the knots. *The absolute value of the observed 
\hb\ flux for the whole nebula was obtained from \citet{b015}. **From the spectra in \pi\ what we actually 
measured was [Ne{\sc iii}]$\lambda$3967.5}+H$\epsilon\lambda$3970.1 and [Ar{\sc iv}]$\lambda$4711+\hei$\lambda$4713. \ \ \ \ \ \ \ \ \ \ \ \ \ \ \ \ \ \ \ \ \ \ \ \ \ \ \ \ \ \ \ \ \ \ \
\end{table*}

\section{Results}
\label{sec:results}

The best fit to the observed spectra was obtained assuming the parameters given
in Table~2, as discussed in Section~\ref{subsec:input}. Table~3 shows the 
predicted and observed intensities of some important collisionally excited 
lines, in which  values are given relative to that of \hb(=100) for each nebular 
component (R and K) and integrated over the nebula (NEB). 

The model \hb\ fluxes are given in the first row of the table. 
The dereddened line intensities quoted in the {\it Obs} column of Table~3 were 
obtained from the spectroscopic data presented in Table~1 of \pi, by using a 
logarithmic extinction constant \chb=0.16 \citep{b017} and the reddening law 
of \citet{b09}. For each line intensity, in each component, the upper row 
(of the {\it Obs} column) 
shows the values for the North-East (R1 and K1) regions of NGC~7009, while those for the South-Western R2 and K4 
components are given in the lower row (see Figure~1 and Table~1 of \pi).

The line intensities predicted by our model were convolved with a long-slit profile (assumed to be rectangular) aligned 
along the long axis of the PN. This is a necessary correction for an extended object with a complex geometry, such as 
that of NGC~7009, if any meaningful conclusions regarding the ionisation and temperature structures are to be gained 
from the comparison of the model with the observations. The dimensions of 1.5\arcsec\ vs. 4\arcmin\ at a distance of 
0.86~kpc were assumed in order to be consistent with \pi. For each nebular
component listed in Table~3 we give the 
{\it slit} and {\it no-slit} line intensities in adjacent columns. 

The absolute value for the observed H$\beta$ flux of NGC~7009 of 3.197 $\times$
10$^{-10}$ erg cm$^{-2}$ s$^{-1}$, quoted in Table~3, was obtained from the VLA
radio recombination line flux \citep{b015}. This flux should be compared to the nebula-integrated 
value (NEB) predicted by our model for the {\it no-slit} case. A very good agreement (better than 2\%) is found.

\subsection{Narrow slit effects}
\label{sub:slitnoslit}

From the comparison of the model {\it slit} and {\it no-slit} columns it appears that some emission lines are more 
affected than others. This is very easy to understand if one considers the electron temperature distribution and 
the physical extension of the various regions where each of the relevant ionic species are most abundant.

He~{\sc ii}$\lambda$4686, for example, is enhanced by $\sim$31\% in the {\it slit} results for the R component and 
$\sim$29\% overall. The opposite behaviour is shown by the He~{\sc i} lines for which the {\it slit} results show a 
$\sim$5\% depletion for He~{\sc i}$\lambda$5876 and He~{\sc i}$\lambda$6678 in both the R component and NEB component. 
Similar beaviours are observed for other species as well, particularly we note the enhancement of intensities in the 
{\it slit} column for [N~{\sc ii}], [S~{\sc ii}] and [S~{\sc iii}] lines. We also note that different lines from the 
same ionic species appear to be enhanced/depleted by different amounts, like [S~{\sc ii}]$\lambda$4069, $\lambda$4076, 
enhanced by $\sim$10-12\% and $\lambda$6717,$\lambda$6731 enhanced by $\sim$15-18\%. This is due to the different 
sensitivities of the various transitions to changes in the electron temperatures.

\subsection{Comparison of the emission lines spectrum}

\begin{table*}
\begin{center}
\caption{Mean temperatures (K) weighted by ionic species.}
\begin{tabular}{lrrrrrrr}
\multicolumn{8}{l}{} \\
\hline
\hline
  	&	&	&	& Ion	 &	&	&	\\
\cline{2-8}
Element	& {\sc i}&{\sc ii}&{\sc iii}&{\sc iv}&{\sc v}&{\sc vi}&{\sc vii}\\
\hline
H	& 10,002& 10,378&	 &	&	&      &       \\
	& 10,625& 10,562& & & & & \\
           	& 10,106& 10,380& & & & & \\
%	&&&&&&&\\
He	&  9,832&  9,955& 12,238&	& 	&	&      \\
	& 10,603& 10,562& 10,565& & & & \\
	&  9,914&  9,968& 12,238& & & & \\
%	&&&&&&&\\
C	&  9,800&  9,874& 10,089& 10,733& 12,887& 12,813& 10,378\\
	& 10,628& 10,598& 10,559& 10,506& 10,494& 10,563& 10,563\\
	& 10,142&  9,961& 10,101& 10,731& 12,887& 12,813& 10,380\\
%	&&&&&&&\\
N	&  9,798&  9,875& 10,098& 10,706& 12,757& 13,158& 13,072\\
	& 10,684& 10,625& 10,555& 10,507& 10,490& 10,563& 10,563\\
	& 10,402& 10,015& 10,110& 10,704& 12,757& 13,158& 13,070\\
%	&&&&&&&\\
O	&  9,795&  9,863& 10,073& 12,452& 12,962& 13,183& 13,301\\
	& 10,697& 10,621& 10,549& 10,515& 10,563& 10,563& 10,563\\
	& 10,529&  9,987& 10,082& 12,452& 12,962& 13,183& 13,301\\
%	&&&&&&&\\
Ne	&  9,805&  9,920& 10,209& 12,590& 13,026& 13,269& 13,312\\
	& 10,627& 10,585& 10,561& 10,528& 10,563& 10,563& 10,563\\
	& 10,041&  9,967& 10,215& 12,590& 13,026& 13,269& 13,312\\
%	&&&&&&&\\
S	&  9,793&  9,845& 10,021& 10,423& 11,341& 13,021& 13,308\\
	& 10,635& 10,605& 10,560& 10,510& 10,467& 10,469& 10,563\\
	& 10,221&  9,973& 10,040& 10,422& 11,339& 13,021& 13,308\\
%	&&&&&&&\\
\hline
\hline
\end{tabular}
\end{center}
\small{For each element, from the upper to the lower row, we show values  
for R, K and NEB, respectively.}
\end{table*}

We show in the last three columns of Table~3 a comparison of our model with the observations integrated over the whole 
slit. Whilst a satisfactory agreement is obtained for many emission lines, some discrepancies do remain, including 
the case of the \nii\ and the \sii\ lines. These discrepancies can be readily understood by noticing the different 
emission lines intensities measured for the North-East and the South-West sides of the nebula (Table~1 of \pi). In our 
models we have assumed the nebula to be symmetric about the x-, y- and z-axis and cannot therefore reproduce such 
asymmetries, which will be reflected into the integrated nebular spectrum for the species that are affected the most. 
This is not a great concern for us, given that a detailed model specific to NGC~7009 is not being sought, our goal 
being the construction of a model able to explain the apparent enhancement of some low-ionisation species in LISs of 
PNe such as NGC~7009 in terms of photoionisation effects only, without the need of assuming enhancements in the total 
elemental abundances of those regions. For this reason, in the modelling we aimed at reproducing the spectra of the 
individual regions (R and K) with values falling between or being close to one of the observational data for North-East 
or the South-West regions. 

A good agreement is shown in the rim and shell (namely R, as justified in Section~2.3.2) with nearly all predicted lines 
falling at intermediate values between the two sides of the nebula (upper and lower rows of the {\it Obs} column), or 
being within 10-30\% of one of the two values.  The only notable exception is the \sii$\lambda\lambda$4068.6, 4076.4 
auroral doublet, with both component being depleted by factors of 4 to 10. The
same behaviour is not observed in the 
knots. We also note that {\sc mocassin} predicts slightly enhanced values for the \sii$\lambda\lambda$6717,6731 nebular 
doublet in the {\it R-component}. Since the auroral lines have much higher critical densities than the nebular lines, it is 
possible that they are being produced in a region that is not accounted for in our model. It would be useful to check 
the behaviour of the \oii$\lambda$7320 doublet, also auroral transitions, but unfortunately these fall outside the 
spectral range explored by \pi. 

The agreement between model and dereddened intensities of the {\it K-component} is even better than that of the 
{\it R-component}. All model predictions fall in between the data observed at the two sides of the nebula, or are within 
10-30\% of one of the two values. Our model does not predict any He~{\sc ii}$\lambda$4686 emission from the knots. We 
argue that the, however quite low, emission reported at the locations of K1 and K4 could be attributed to intervening 
material in the PN halo not included in our model. 

In the following we will discuss in more detail the predicted mean temperatures and ionic 
abundances derived from our modelling.

\subsection{Mean temperatures}

Mean temperatures weighted by the ionic abundances are given in Table~4. Expressions for the volume integrated 
fractional ionic abundances (discussed in Section~3.4) as well as those for the mean temperatures weighted by 
ionic abundances were defined in \citet{b011}, their eqs. (2) and (3). The 
values listed in Table~4 show that the 
temperature structure of the nebula is well reproduced by our model. Figures of \te\oiii\ obtained for regions 
R, K and NEB are within the empirical values determined in \pi\ (see Table~1) for the two sides of the nebula, namely 10,100~K, 
10,000~K, and 10,100K for rim, knots and NEB, respectively. Similarly, the values of \te\nii\ of the three 
components agree with the empirical values to better than 5\%, while for \te\sii\ the agreement is to within 33\% and 
11\% for the two knots, due to our model underestimating of the \sii\ auroral lines (see Section~3.2). 

We confirm that the electron temperature distribution is fairly homogeneous across the various regions of the nebula,  
see for instance, \citet{b038} and \citet{b039}. The mean \te\ for the whole nebula, from the neutral (I) to the highly 
ionized ions (VII), for all species, are respectively 10,200$\pm$200K, 10,050$\pm$150K, 10,050$\pm$850K, 11,400$\pm$1050K, 
12,600$\pm$700K, 13,100$\pm$200K and 12,650$\pm$1,300K.

Our model results for \ne\ and \te\ compare well with the many estimates available from the literature; in particular 
\te\oiii\ and \te\nii\, for the rim, can be compared to the values of \citet{b022,b023,b031,b026,b029,b039} that cluster 
around 10,000K for \te\oiii, while values between 9,000K and 12,800K, peaking around 10,550K, were found for \te\nii\ 
by \citet{b022,b023,b026,b017}. As for the outer knots, \te\nii\ and \te\oiii\ were found as being 8,100K and 11,500K 
by \citet{b06}, 9,300K and 9,100K by \citet{b026} and 11,350 and 10,000K by \citet{b017}.

\subsection{Fractional ionic abundances}

\begin{table*}
\begin{center}
\caption{Averaged fractional ionic abundances.}
\begin{tabular}{llllllll}
\multicolumn{8}{l}{} \\
\hline
\hline
  	&	&	&	& Ion	 &	&	&	\\
\cline{2-8}
Element	& {\sc i}&{\sc ii}&{\sc iii}&{\sc iv}&{\sc v}&{\sc vi}&{\sc vii}\\
\hline
H	& 5.40(-4)& 0.999   & 	      & 	& 	  & 	    & 	      \\
	& 1.04(-2)& 0.989   &         &         &         &         &         \\
	& 6.46(-4)& 0.999   &&&&& \\
%	&&&&&&&\\
He	& 7.00(-4)& 0.814   & 0.185   &  	& 	  & 	    & 	      \\
	& 7.39(-3)& 0.992   &         &         &         &         &         \\
	& 7.74(-4)& 0.817   & 0.181&&&&\\
%	&&&&&&&\\
C	& 3.10(-6)& 1.08(-2)& 0.555   & 0.428	& 5.45(-3)&	    &	      \\
	& 2.42(-4)& 0.132   & 0.834   & 3.24(-2)&         &         &         \\
	& 5.33(-6)& 1.21(-2)& 0.560   & 0.421   & 5.33(-3)&         &\\
%	&&&&&&&\\
N	& 1.37(-6)& 5.84(-3)& 0.546   & 0.443	& 4.12(-3)& 1.00(-5)& 	      \\
	& 3.73(-4)& 0.136   & 0.823   & 4.01(-2)&         &         &         \\
	& 4.47(-6)& 7.14(-3)& 0.551   & 0.437   & 4.03(-3)& 9.85(-6)&         \\
%	&&&&&&&\\
O	& 4.83(-6)& 9.72(-3)& 0.862   & 0.125	& 3.13(-3)& 1.11(-6)&         \\
	& 2.86(-3)& 0.188   & 0.808   &         &         &         &         \\
	& 2.81(-5)& 1.15(-2)& 0.863   & 0.122   & 3.07(-3)& 1.08(-3)& 1.08(-6)\\
%	&&&&&&&\\
Ne	& 3.56(-6)& 1.07(-2)& 0.917   & 7.11(-2)& 7.20(-4)&         &         \\
	& 1.56(-4)& 6.67(-2)& 0.933   &         &         &         &         \\
	& 4.98(-6)& 1.14(-2)& 0.918   & 6.95(-2)& 7.04(-4)&         &\\
%	&&&&&&&\\
S	& 6.84(-7)& 8.13(-3) & 0.343   & 0.546	& 9.89(-2)& 2.64(-3)& 9.97(-6)\\
	& 8.16(-5)& 0.155   & 0.755   & 8.81(-2)& 6.71(-4)&         &         \\
	& 1.42(-6)& 9.71(-3)& 0.350   & 0.540   & 9.70(-2)& 2.58(-3)& 9.75(-6)\\
%	&&&&&&&\\
Ar      & 1.52(-7)& 1.25(-3)& 0.363 & 0.613 & 1.90(-2) & 2.86(-3) & 2.51(-5)\\
        & 2.99(-5)& 2.20(-2)& 0.821 & 0.156 &  & & \\
	& 4.05(-7)& 1.47(-3)& 0.370 & 0.606 & 1.85(-2) & 2.80(-3) & 2.46(-5)\\
\hline
\hline
\end{tabular}
\end{center}
\small{For each element the first row is for the R, the second  
is for K, and the last one is for NEB.}
\end{table*}

\begin{figure*}
\psfig{file=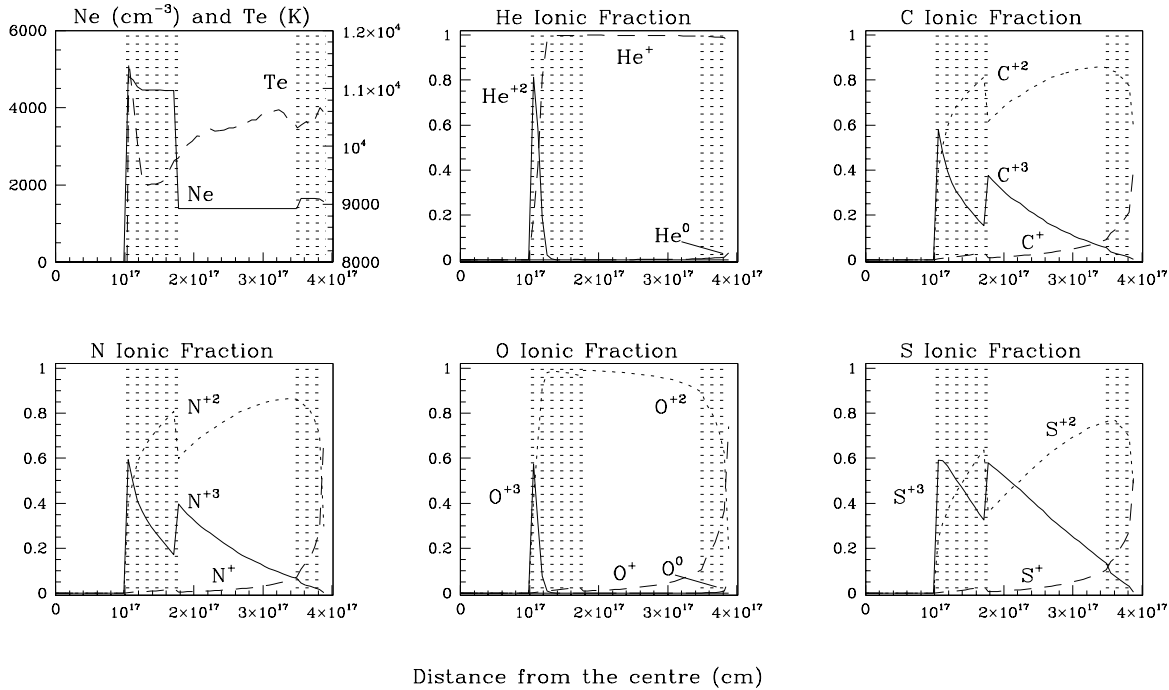,width=16.0truecm,bbllx=35pt,bblly=280pt,bburx=570pt,bbury=600pt}
 \caption{Model profiles along the major axis. From left to right and top to bottom, the electron density and 
temperature, as well the ionic fraction for He, C, N, O and S are shown. The position of the {\it R-component} 
(9.9$\times$10$^{16}$ up to 1.84$\times$10$^{17}$cm) and that of the {\it K-component} (3.49$\times$10$^{17}$ up 
to 3.87$\times$10$^{17}$cm) are marked in each panel. After the inner cavity the inner wall of the ellipsoidal 
shell is encountered, this has N$_H$ = 4,000~cm$^{-3}$ and N$_e$ $\sim$ 5,000~cm$^{-3}$. The value of N$_e$ decreases 
through the first shaded area that represents the ellipsoidal shell, as the level of ionisation of the gas decreases. 
After the first shaded area we find the jet (N$_H$=1,250~cm$^{-3}$, N$_e$ $\sim$ 1,500~cm$^{-3}$). At the end of the 
jet we find the knot, where and the N$_H$ = 1,500~cm$^{-3}$, while N$_e$ reaches about 1,700~cm$^{-3}$.}
\label{IonProf}
\end{figure*}

Results for the fractional ionic abundances are shown in Table~5. Hydrogen and helium are fully at least singly-ionized in R, K 
and NEB, and significant fractions of the heavy elements are in higher ionization stages in the rim as well as in 
the nebula as a whole (NEB). We also note the lower ionization of the knots. An important issue that should be 
noted here is the N/N$^+$ ratio being higher than the O/O$^+$ ratio by a factor of 1.39 in the knots, 1.66 rim 
and 1.61 in the total nebula. This result is at variance with the N/N$^+$=O/O$^+$ \citep{b025,b034} generally 
assumed by the {\it icf} method, with the consequent errors on empirically derived total elemental abundances, 
like those in \pi.
 
We note from Table~5, that only a small fraction ( 0.6\%, 14\% and 0.7\% for R, K and NEB, respectively) of the total 
nitrogen in the nebula is in the form of N$^0$ and N$^+$. As only lines from these ions were observed 
(see Table~3 of \pi), the nitrogen abundance determination is particularly uncertain. Sulfur abundances also suffer 
from similar problems. 

Combining the values in Table~5 with those for the total elemental abundances 
used by our model (Table~2), 
fractional ionic abundances relative to H are readily available. The total abundance that we have used 
for all regions modeled returns ionic abundances, for R and K respectively, of He$^+$~=~0.091 and 0.111;  
N$^+$~=~1.17$\times$10$^{-6}$ and 2.72$\times$10$^{-6}$;  O$^{+2}$~=~3.88$\times$10$^{-4}$ and 
3.64$\times$10$^{-4}$; Ne$^{+2}$~=~9.72$\times$10$^{-5}$ and 9.89$\times$10$^{-5}$; and finally  
S$^{+2}$~=~3.09$\times$10$^{-6}$ and 6.79$\times$10$^{-6}$. These ---which are the most important 
ions of the optical range for determining the total abundances--- are comparable to the ionic abundances 
derived from the observations in \pi, with an agreement better than 90\% for all of the above ions, 
except the S$^{+2}$/H of R, whose discrepancy reaches 70\%. 

In Figure~\ref{IonProf} we show the ionic abundance profiles of the model, as well as the electron density and 
temperature profiles along the axis which includes the knots. The top-left panel in this Figure, reflects the 
\ne\ (continuous line) and \te\ (dashed line) variations through the long axis of the nebula. It shows that \ne\ 
peaks at the innermost region of the {\it R-component}, having a mean value through the component of 4,100~cm$^{-3}$. 
The {\it J-component} as well as the {\it K-component} have mean \ne\ of 1385~cm$^{-3}$ and 1640~cm$^{-3}$, respectively, 
while somewhat smoothed profiles are present at the edges. The electron density profile is a result of the ionization 
structure combined with the density 
distribution assumed in Section~2.3.3. As for the \te\ profiles, although varying somewhat within the components, have 
mean values that are in close agreement with the values measured in \pi, see numbers quoted in the previous 
subsection. This will be discussed further in Section~4.

Figure~1 also highlights the strong dependence of the 
ionisation level on the geometry and density distribution of the gas. It is therefore clear that an apparent 
overabundance of N$^+$ in the knots can be produced by providing the correct gas opacity to screen this region 
from the direct stellar photons. We should add at this point that an even larger N$^+$ abundance can be obtained 
by further enhancing the gas density at the rim-jet interface, without significantly changing the [S~{\sc ii}] 
density ratio in the {\it R-component}. 

\subsection{Visualization of the model results}

\begin{figure*}
\psfig{file=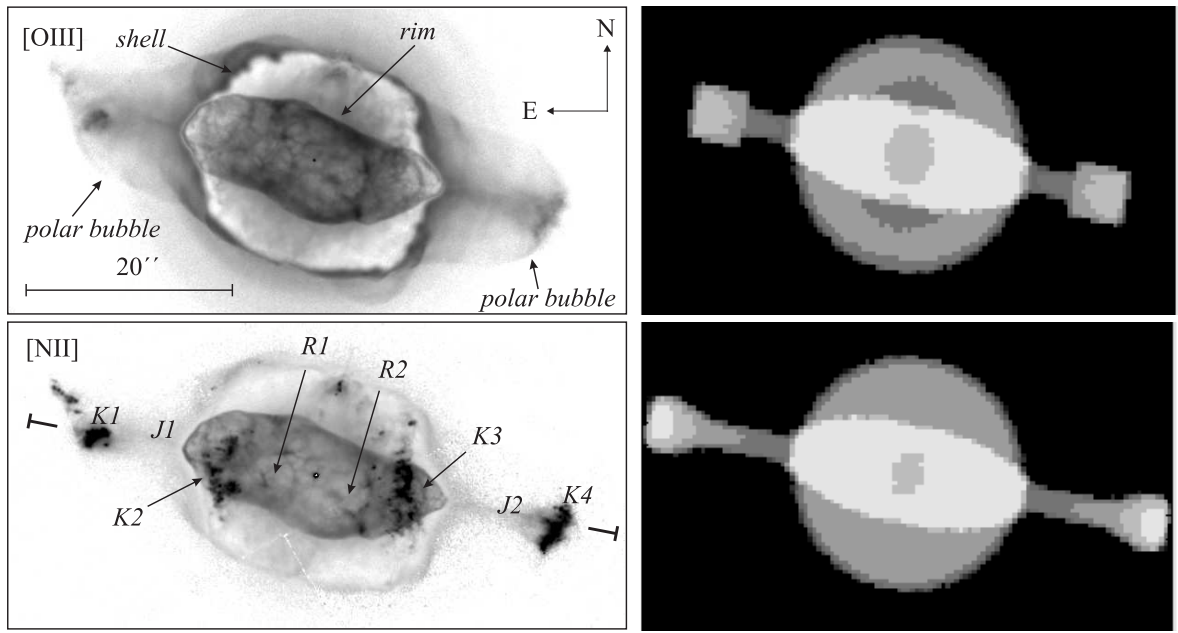,width=16.0truecm,bbllx=0pt,bblly=0pt,bburx=630pt,bbury=330pt}
\caption{Left panel: {\itshape HST} \oiii\ and \nii\ images of NGC~7009 on a logarithmic 
intensity scale; slit position for the spectra discussed in Section~3 is indicated by short lines 
(P.A.  = 79$^\circ$), while labels mark the position of the several structures, where ``K'', ``J'' 
and ``R'' stand for ``knot'', ``jet'', and ``rim'', respectively. Right panel: project emission maps 
from the model; the top map is \oiii\ and the bottom one is \nii; maps (in arbitrary units) are not 
on the same colour intensity scale.}
\label{EmisMap}
\end{figure*}

NGC~7009 has been observed with the {\it HST} WFPC2 with filters centred in the \oiii$\lambda$5007 and 
\nii$\lambda$6583 emission lines (see Section~2.1). These archive images, already published in \pi, are 
compared to the model predicted emission maps in Figure~\ref{EmisMap}. The maps were 
produced at an inclination of 84$^\circ$ with respect to the line of sight, as indicated by the 
kinematics of the PN polar axis \citep{b037}, for \oiii\ (right top panel) and  
\nii\ (lower right panel) emission lines. First of all we call the readers attention 
to the fact that the {\it polar bubble}, which appear in the {\it HST} \oiii\ image, at fainter intensity 
levels, as an extension of the {\it shell}, and also the inner pair of knots, {\it K2} and {\it K3} 
were not considered in our modelling. Thus, excluding the polar bubble and the inner 
knots, the images and our maps are at least qualitatively in good agreement (note that a photometric comparison 
is beyond the scope of this paper). 

From the maps we note the higher excitation of the equatorial axis of the PN as 
compared to the polar one, because of the higher density of the rim, and following the pattern known 
from previous studies, as stated in the Introduction. One also clearly sees that the \nii\ map is more 
extended than the \oiii\ one, as expected from a nebula excited by a central star. Finally, one notes 
that the knots are fainter (as compared to the inner regions) in \oiii\ and brighter in \nii, due to the 
enhanced recombination in the knots. All the above suggest that our model can satisfactorily reproduce the 
main features in the images of NGC~7009, that could in principle be extended to other PNe with a pair of 
outer low-ionization knots \citep[see, for more examples,][]{b016}.

\section{Alternative Model}

As noted in previous sections, the ionisation structure of the model is 
strongly dependent on the input 3D density distribution. Because of that, here we also present 
the results obtained with a slightly different geometry, which we call the {\it alternative model}. 
The only difference in terms of input density distribution between our initial model and the alternative 
one is that in the latter the jets start 3.3\% further away from the central star than in the previous 
model.  Since the H number density in the jet is four times lower than the value in the rim at the position 
of the polar axis, our change results into an increase in the line optical depth along the path  
connecting the central star to the outer pair of knots. The intensity of the ionizing stellar field reaching 
the knots is therefore decreased in our alternative model, hence resulting in a lower ionization level. Other 
model parameters were also adjusted in order to optimise our fit of the spectrum and these are summarised 
in Table~6. More importantly, with this alternative model we aimed at reproducing the more extreme \nii\ 
emission of the Eastern knot of the nebula (\ki\ in \pi\ and in Figure~2).
\begin{table}
\begin{center}
\caption{Input parameters for the alternative model.}
\begin{tabular}{llll}
\hline
\hline
\multicolumn{4}{l}{} \\           
L$_*$ ($L_{\odot}$) & 3136                  & N/H    & 2.8$\times$10$^{-4}$  \\
T$_{eff}$ (K)       & 82,000 	            & O/H    & 4.5$\times$10$^{-4}$  \\
R$_{in}$ (cm)       & 0.0 	            & Ne/H   & 1.0$\times$10$^{-4}$ \\
R$_{out}$ (cm)      & 3.88$\times$10$^{17}$ & S/H    & 0.9$\times$10$^{-5}$  \\
He/H                & 0.12                  & Ar/H   & 1.5$\times$10$^{-6}$  \\
C/H                 & 3.5$\times$10$^{-4}$  & Fe/H   & 1.25$\times$10$^{-6}$  \\
\hline
\hline
\end{tabular}
\end{center}
\small{Abundances are given by number, relative to H.} 
\end{table}

\begin{table*}
\begin{center}
  \caption{Alternative model and observed dereddened spectra of R, K and NEB. Line intensities are normalized to \hb=100.} 
\begin{tabular}{llllllllll}
\hline
\hline
\noalign{\smallskip}
Line Identification (\an) & \multicolumn{3}{c}{R} & \multicolumn{3}{c}{K} & \multicolumn{3}{c}{NEB}\\
       & Model     & Model  & Obs. &  Model	 & Model  & Obs. & Model	 & Model  & Obs.  \\
       & no-slit & slit &      & no-slit & slit &      & no-slit & slit	&       \\
\noalign{\smallskip}
\hline
\multicolumn{10}{l}{}\\
\hb (10$^{-13}$ erg cm$^{-2}$ s$^{-1}$)& 3120 & 398.5 & - &  7.95 & 2.95 & - & 3137 & 407.1 & 3197* \\ \\
\hline
{}[O{\sc ii}] 3726.0 + 3728.8 & 5.37 & 5.95 & 15.5 &  257. & 299. & 204. & 6.18 & 8.81 & 24.2\\
                              &      &      & 7.34 & 	   &	  & 155. &	&      &     \\
{}[Ne{\sc iii}] 3868.7        & 109. & 110. & 106. &  123. & 122. & 82.3 & 109. & 110. & 108.\\
                              &      &      & 105. & 	   &	  & 141. &	&      &     \\
{}**[Ne{\sc iii}] 3967.5      & 33.8 & 34.0 & 51.7 &  38.1 & 37.9 & 35.3 & 33.8 & 34.1 & 48.9\\
                              &      &      & 44.6 & 	   &	  & 59.8 &	&      &     \\
{}[S{\sc ii}] 4068.6          & 0.32 & 0.36 & 1.50 &  5.89 & 6.72 & 7.43 & 0.33 & 0.42 & 1.98\\
                              &      &      & 2.36 & 	   &	  & 5.04 &	&      &     \\
{}[S{\sc ii}] 4076.4          & 0.10 & 0.12 & 0.87 &  1.95 & 2.22 & 2.36 & 0.11 & 0.14 & 1.11\\
                              &      &      & 1.16 & 	   &	  & 1.83 &	&      &     \\
{}[O{\sc iii}] 4363.2         & 9.11 & 9.13 & 7.76 &  9.86 & 9.39 & 6.94 & 9.12 & 9.16 & 8.15\\
                              &      &      & 8.65 & 	   &	  & 9.64 &	&      &     \\
He{\sc ii} 4685.7             & 18.4 & 24.0 & 26.0 &  0.00 & 0.00 & 1.00 & 18.3 & 23.5 & 15.8\\ 
                              &      &      & 23.4 & 	   &	  & 1.23 &	&      &     \\
{}**[Ar{\sc iv}] 4711         & 4.28 & 4.23 & 5.48 &  1.49 & 1.38 & 2.19 & 4.27 & 4.19 & 4.10\\
                              &      &      & 4.96 & 	   &	  & 2.00 &	&      &     \\
{}[Ar{\sc iv}] 4740.2         & 5.67 & 5.73 & 5.56 &  1.26 & 1.17 & 1.14 & 5.65 & 5.65 & 3.92\\
                              &      &      & 4.54 & 	   &	  & 1.49 &	&      &     \\
H$\beta$ 4861.3	              & 100. & 100. & 100. &  100. & 100. & 100. & 100. & 100. & 100.\\
                              &      &      & 100. & 	   &	  & 100. &	&      &     \\
{}[O{\sc iii}] 5006.8         & 1234 & 1208 & 1162 &  1262 & 1205 & 1238 & 1234 & 1212 & 1206\\
                              &      &      & 1225 & 	   &	  & 1310 &	&      &     \\
{}[Cl{\sc iii}] 5517.7        & 0.35 & 0.36 & 0.43 &  1.07 & 1.05 & 0.00 & 0.35 & 0.37 & 0.54\\
                              &      &      & 0.43 & 	   &	  & 0.97 &	&      &     \\
{}[Cl{\sc iii}] 5537.9        & 0.57 & 0.59 & 0.53 &  1.03 & 1.01 & 0.00 & 0.57 & 0.60 & 0.64\\
                              &      &      & 0.55 & 	   &	  & 0.90 &	&      &     \\
{}[N{\sc ii}] 5754.6          & 0.15 & 0.16 & 0.14 &  5.33 & 6.32 & 6.49 & 0.16 & 0.22& 0.46\\
                              &      &      & 0.18 & 	   &	  & 3.91 &	&      &     \\
He{\sc i} 5875.7              & 15.0 & 14.3 & 13.9 &  17.0 & 17.0 & 18.8 & 15.0 & 14.4 & 14.5\\
                              &      &      & 14.1 & 	   &	  & 15.0 &	&      &     \\
{}[S{\sc iii}] 6312.1         & 1.50 & 1.57 & 1.27 &  3.97 & 3.91 & 3.89 & 1.50 & 1.61 & 1.68\\
                              &      &      & 1.14 & 	   &	  & 3.25 &	&      &     \\
{}[N{\sc ii}] 6583.4          & 8.12 & 9.32 & 7.09 &  296. & 352. & 355. & 9.03 & 12.5 & 27.\\
                              &      &      & 5.60 & 	   &	  & 194. &	&      &     \\
He{\sc i} 6678.1              & 4.25 & 4.06 & 3.87 &  4.81 & 4.82 & 7.49 & 4.26 & 4.08 & 3.97\\
                              &      &      & 3.95 & 	   &	  & 3.18 &	&      &     \\
{}[S{\sc ii}] 6716.5          & 0.45 & 0.54 & 0.48 &  26.0 & 29.8 & 36.8 & 0.54 & 0.85 & 2.33\\
                              &      &      & 0.38 & 	   &	  & 23.2 &	&      &     \\
{}[S{\sc ii}] 6730.8          & 0.84 & 1.00 & 0.84 &  34.9 & 39.9 & 51.0 & 0.96 & 1.41 & 3.85\\
                              &      &      & 0.69 & 	   &	  & 28.4 &	&      &     \\
\noalign{\smallskip}
\hline
\hline
\end{tabular}
\end{center}
\small{As in Table~2, with the top/bottom rows given intensities of the Eastern/Western size of the nebula.}
\end{table*}

\begin{table*}
\begin{center}
\caption{Averaged fractional ionic abundances for the alternative model.}
\begin{tabular}{llllllll}
\multicolumn{8}{l}{} \\
\hline
\hline
  	&	&	&	& Ion	 &	&	&	\\
\cline{2-8}
Element	& {\sc i}&{\sc ii}&{\sc iii}&{\sc iv}&{\sc v}&{\sc vi}&{\sc vii}\\
\hline
H	& 5.53(-4)& 0.999   & 	      & 	& 	  & 	    & 	      \\
	& 1.38(-2)& 0.986   &         &         &         &         &         \\
	& 6.91(-4)& 0.999   & 	      & 	& 	  & 	    & 	      \\
%	&&&&&&&\\
He	& 6.95(-4)& 0.809   & 0.190   &  	& 	  & 	    & 	      \\
	& 8.99(-3)& 0.991   &         &         &         &         &         \\
	& 7.83(-4)& 0.813   & 0.186   &  	& 	  & 	    & 	      \\
%	&&&&&&&\\
C	& 3.19(-6)& 1.06(-2)& 0.542  & 0.440	& 6.48(-3)&	    &	      \\
	& 3.27(-4)& 0.155   & 0.817  & 2.65(-2) &         &         &         \\
	& 6.13(-6)& 1.21(-2)& 0.547  & 0.433	& 6.35(-3)&	    &	      \\
%	&&&&&&&\\
N	& 1.39(-6)& 5.70(-3)& 0.533   & 0.455	& 5.08(-3)& 1.62(-5)& 	      \\
	& 6.83(-4)& 0.176   & 0.789   & 3.31(-2)&         &         &         \\
	& 6.86(-6)& 7.34(-3)& 0.539   & 0.448	& 4.972(-3)& 1.58(-5)& 	      \\
%	&&&&&&&\\
O	& 4.90(-6)& 9.40(-3)& 0.856   & 0.130	& 3.94(-3)& 1.95(-6)&         \\
	& 5.29(-3)& 0.235   & 0.759   &         &         &         &         \\
	& 4.68(-6)& 1.15(-2)& 0.856   & 0.127	& 3.85(-3)& 1.90(-6)&         \\
%	&&&&&&&\\
Ne	& 3.42(-6)& 1.02(-2)& 0.910   & 7.85(-2)& 1.03(-3)&         &         \\
	& 2.09(-4)& 7.17(-2)& 0.928   &         &         &         &         \\
	& 5.27(-6)& 1.09(-2)& 0.911   & 7.69(-2)& 1.00(-3)&         &         \\
%	&&&&&&&\\
S	& 7.12(-7)& 8.03(-3)& 0.333   & 0.548	& 0.105  & 3.38(-3)& 1.58(-5)\\
	& 1.13(-4)& 0.190   & 0.734   & 7.43(-2)& 5.21(-4)&         &         \\
	& 1.72(-6)& 9.94(-3)& 0.340   & 0.542	& 0.103  & 3.31(-3)& 1.55(-5)\\
%	&&&&&&&\\
Ar      & 1.54(-7)& 1.22(-3)& 0.351 & 0.621 & 2.17(-2) & 3.94(-3) & 4.39(-5)\\
        & 5.51(-5)& 2.74(-2)& 0.825 & 0.147 &  & & \\
	& 6.02(-7)& 1.48(-3)& 0.358 & 0.614 & 2.12(-2) & 3.86(-3) & 4.30(-5)\\
	
\hline
\hline
\end{tabular}
\end{center}
\small{Top to bottom rows give values for R, K and NEB, respectively.}
\end{table*}

As in the case of our original model discussed in Section~3, we found a good agreement of the alternative model 
with the dereddened fluxes of \pi, for the rim, the knots as well as for the whole nebula (NEB). Again, most 
of the line intensities are within 10-30\% of one of the sides of NGC~7009, or between the two values of the Eastern 
and Western sides, as quoted in Table~7, with the exceptions of the \sii$\lambda\lambda$4068,4076 and 
\sii$\lambda\lambda$6717,6731 doublets, as pointed out in Section~3.2. 

The mean \te\ of the alternative model also compare very nicely with the observed values, namely, 
\te\oiii\ and \te\nii\ of 10,200~K and 9,950~K, 10,600~K and 10,600~K, and 10,200~K and  10,100~K, 
for rim, knots and NEB, respectively. And the \ne\ are automatically matched with the empirical values, 
since they are constraining the model as input parameters. 

The corresponding ionisation structure is shown in Table~8, from which we obtain the N/N$^+$ 
and O/O$^+$ ratios of the different zones in the nebula, as being: 1.64 ({\it R-component}), 1.33 
({\it K-component}) and 1.56 (NEB), that, as above, are in contradiction with the ratios adopted in 
the empirical {\it icf} scheme. We chose to include Table~8 in the paper, although qualitatively 
similar to Table~5, to allow for model {\it icf}s to be derived at a later time should one wish to 
do so. 

Finally, the projected emission maps obtained from this model are not shown here as they are qualitatively 
identical to those of our original model (right panels of Figure~2).

\section{Conclusions}

This work focused on the study of the apparent N overabundance in the outer knots of NGC~7009 with respect to 
the main nebular rim and shell. 

\citet{b01} and \citet{b020} showed that long-slit data may give spurious overabundances of N and other 
elements in the outer regions of model PNe. These authors have identified the different charge-exchange 
reaction rates of N and O as the main responsible for the effect in the low-ionization regions of the nebulae,  
therefore affecting the (N$^+$/N)/(O$^+$/O) ratio, that as discussed in Section~3.4, is commonly used to obtain 
the {\it icf} for nitrogen. However, as shown by \citet{b030} charge-exchange reactions can, in this case, 
account at most by 20\% of the nitrogen overabundance of the knots of NGC~7009. 

In this paper we have presented a model that was able to reproduce the main spectroscopic characteristics  of the 
various spatial regions of NGC~7009 without the need of assuming an inhomogeneous set of abundances.
We investigated the importance of taking into account the effects of a narrow 
slit, and our results in Table~3 and 7 
show 
that the convolution of the model results with the profile of the narrow slit used for the observations presented 
in \pi\ results in the [N~{\sc ii}] emission being enhanced respect to \hb\ in all regions of the nebula, with the 
effect being slightly more pronounced in the knots. 

The (N$^+$/N)/(O$^+$/O) predicted by our models are 0.60, 0.72 and 0.62 (or alternatively, 0.61, 0.75 and 0.64) 
for the rim, knots and the whole nebula, respectively, 
all at variance with the {\it icf} assumption of unity for this ratio. The {\it icf}s will therefore be underestimated 
by the empirical scheme, both in the case of the R and K components, but more so in the former (by a factor of 1.21). 
Therefore this effect may partly be responsible for the apparent inhomogeneous N abundance derived from observations. 
The differences in the (N$^+$/N)/(O$^+$/O) in these two components may be due to a number of effects, including charge 
exchange, as stated above, and the difference in the ionization potentials of the relevant species, which makes the 
(N$^+$/N)/(O$^+$/O) ratio extremely sensitive to the shape of the local radiation field. For this reason, a realistic 
density distribution is essential to the modelling of a non-spherical object, if useful information is to be extracted 
from spatially resolved observations. The density distribution of the gas modifies the shape of the local radiation 
field via the gas opacities and (as demonstrated in Section~3.4 and Section~4) matching the emission line spectrum of a 
given nebular region relies, among other things, on the careful construction of a model capable of providing the correct 
optical depths in various directions. 

Our main conclusions discussed here may also be extended to other PNe exhibiting FLIERs, such as NGC~6543 and NGC~6826, 
for which similar apparent chemical inhomogeneities have been claimed \citep{b06}.

\section*{Acknowledgments}

We acknowledge the inspiring ideas of B. Balick on which this work is based, 
and the anonymous referee for his/her positive and helpful comments. We thank 
the Instituto de Astrof\'\i sica de Canarias for the use of the PC cluster 
``beoiac". The work of DRG is supported by the Brazilian Agency FAPESP 
(04/11837-0). BE would like to thank the FAPESP for the visiting grant 
(03/09692-0). We also acknowledge the partial support of the Spanish Ministry 
of Science and Technology (AYA 2002-0883).

\bsp

\label{lastpage}

\end{document}